\documentclass[11pt]{article}
\usepackage{rotating}
\usepackage{amsmath, latexsym}
\usepackage{graphicx}
\usepackage{amsfonts}
\usepackage{amssymb}
\usepackage{subfig}
\usepackage{setspace}
\usepackage{epstopdf}
\usepackage[hidelinks]{hyperref} 
\usepackage[margin=1in]{geometry}
\usepackage{pdflscape}
\usepackage{color}
\usepackage{placeins}
\usepackage{natbib}
\usepackage{afterpage}

\newcommand{\calS}{{\cal S}}
\newcommand{\calT}{{\cal T}}

\newcommand{\given}{\,|\,}

\graphicspath{ {/Users/n_a_abdallah/Desktop/spatial/Project1} }
\usepackage{graphicx}
\usepackage{lipsum}
\usepackage{mathtools}

\usepackage{listings}
\begin{document}

\begin{center}
{\Large \textbf{Coastline Kriging: A Bayesian Approach} }

\bigskip

\textsc{Nada Abdalla$^{1}$, Sudipto Banerjee$^{1*}$, Gurumurthy Ramachandran$^{2}$, Mark Stenzel$^{3}$ and Patricia A. Stewart$^{4 }$}\\
$^{1}$ Department of Biostatistics, University of California-Los Angeles, Los Angeles, California.\\
$^{2}$ Department of Environmental Health and Engineering, Bloomberg School of Public Health, Johns Hopkins University, Baltimore, MD 21205.\\
$^{3}$ Exposure Assessment Applications, LLC.\\
$^{4}$ Stewart Exposure Assessments LLC.\\
$^{1}$ \emph{email:} nada.a.abdallah@gmail.com\\
\end{center}
\textsc{Abstract}
Statistical interpolation of chemical concentrations at new locations is an important step in assessing a worker's exposure level. When measurements are available from coastlines, as is the case in coastal clean-up operations in oil spills, one may need a mechanism to carry out spatial interpolation at new locations along the coast. In this paper we present a simple model for analyzing spatial data that is observed over a coastline. We demonstrate four different models using two different representations of the coast using curves. The four models were demonstrated on simulated data and one of them was also demonstrated on a dataset from the GuLF STUDY. Our contribution here is to offer practicing hygienists and exposure assessors with a simple and easy method to implement Bayesian hierarchical models for analyzing and interpolating coastal chemical concentrations.


\textsc{KEYWORDS} Gaussian process; Hierarchical modeling; Kriging; Markov chain Monte Carlo; Coastal kriging; Geostatistics
\newpage
\section*{Introduction}\label{sec:intro}
Data observed over locations with known geographic coordinates are often referred to as point-referenced data and are commonly seen in environmental health. Recent applications consider such data measured along coastlines or shores. For example, assessing exposures of workers to chemicals along the coastline may require statistical interpolation of the chemical concentration at unmonitored locations along the coast.
Statistical interpolation at new locations based upon a set of observed measurements at known locations is often referred to as  ``Kriging'' in the geostatistical literature \citep{cressie}. Kriging customarily uses spatial analytic tools such as variograms or covariance functions to construct best linear unbiased predictors. When chemicals are sampled mostly along a coastline, interpolation is sought at new locations along the coast. Thus, all measurements are collected along a curve (approximating the coastline) and prediction is sought at new points on this curve. We call this ``coastal kriging.'' 

Models for waterway stream networks using moving averages have been developed \citep{networks}. They use stream distance rather than Euclidean distance. These models account for the volume and direction of flowing water in stream networks. They offer richness and flexibility, but are complicated and can be difficult to compute. Unlike networks, where we have complex structure of line-segments and joints, in simple coastal kriging we approximate the coastline with a single curve or a sequence of line segments. A simple parametrization of the coast will suffice and lead to easily implementable statistical models.

We will pursue Bayesian coastal kriging. Bayesian models offer easier interpretability for parameter estimates, provide exact estimates of uncertainty without requiring assumptions of large sample sizes and independence of observations, and can incorporate prior information when available. Incorporating prior information is not uncommon in exposure assessment and can improve decision making \citep[see, e.g.,][]{rambanvin,hewett}. Bayesian models can be easily executed using several software packages within the \texttt{R} statistical computing environment (please refer to the coding material as seen in the Online Supplementary Material). 

We will illustrate our models using a specific dataset extracted from the GuLF STUDY (Gulf Long-term Follow-up Study) database. In April, 2010 an explosion of the \textit{Deepwater Horizon} oil rig resulted in an oil spill in the Gulf of Mexico. It was the largest oil spill in US history.  Tens of thousands of workers were involved in stopping and in cleaning up the oil release. The GuLF STUDY is conducted by the National Institute of Environmental Health Sciences (NIEHS) and sponsored by the National Institute of Health (NIH) \citep{overview}. It is collecting information to study potential adverse effects on the health of those workers. Among other activities, the workers capped the well, applied dispersants to break up the oil, skimmed or burned the oil on the Gulf waters, cleaned beaches, marshes and structures, decontaminated equipment, and provided support for these activities. Personal air measurements are available on many of these tasks. The highest portion of the STUDY participants were involved in cleaning the beaches, marshes and structures. One specific task in assessing exposures of workers cleaning the coastline is to statistically interpolate the chemical concentration at new locations along the coast. 

Our contribution in this article expands upon existing geostatistical models to allow for better prediction of quantities of interest at new locations over coastlines. The article is organized as follows. Section~2 provides a brief review of Bayesian methods for kriging. Section~3 discusses spatial processes for coastline measurements. Section~4 discusses our geostatistical models for interpolating point-referenced coastline data and simple algorithms for implementing Bayesian kriging. Section~5 discusses simulation results that help validate our method. Section~6 illustrates our model through applying it to the GuLF STUDY data. Section~7 concludes the article suggesting some future work.

\section*{Model-based Kriging}\label{sec:review}
\subsection*{Spatial process models with Euclidean coordinates}\label{sec:spatial_process_euclidean}

Point-referenced spatial modeling seeks to capture associations between observations geographically closer to each other and to predict the value of the response or outcome variable at arbitrary locations. This is achieved using a spatial regression model,
\begin{equation} \label{eq:spatial_basic}
Z(s)=x(s)^{\top}\beta + \omega (s) +\epsilon (s)\; ,\; \epsilon(s) \stackrel{iid}{\sim} N(0,\tau^2)\;,
\end{equation}
where $x(s)^{\top}$ is a $1\times p$ vector of covariates (predictors) observed at location $s$, $\omega(s)$ is a latent (unobserved) spatial random effect at location $s$, and $\epsilon(s)$ accounts for measurement error. For any collections of locations, the measurement errors in (\ref{eq:spatial_basic}) are normally distributed independently and identically, each with a zero mean and variance $\tau^2$. 

If $\omega(s) = 0$ for all locations, then (\ref{eq:spatial_basic}) reduces to an ordinary linear model with independent outcomes. If the outcomes are spatially correlated, then $\omega(s)$ introduces dependence. There are several different mechanisms for specifying $\omega(s)$ \citep{cressie, banerjee}, but we choose a fairly straightforward and interpretable model here. We assume that each $\omega(s)$ has mean $0$ and the dependence at two points $s$ and $s'$ is modeled as
\begin {equation} \label{eq:cov_fun} 
\mbox{Cov}\{\omega(s), \omega(s')\} = K_{\theta}(s,s') = \sigma^2 \exp(-\phi \|s-s'\|)\; ,
\end{equation}
where $\|s-s'\|$ is the distance between two locations $s$ and $s'$, $\sigma^2$ captures the variation attributed to spatial effects (referred to as partial sill) and $\phi$ controls the rate at which the spatial correlation drops to zero. The process parameters $\theta = \{\sigma^2, \phi\}$ together with the distance between any two points completely specify the spatial covariance function $K_{\theta}(s,s')$ The \emph{spatial range} is defined as the distance beyond which the spatial correlation becomes negligible. For the exponential covariance function in (\ref{eq:cov_fun}), the spatial range is given by approximately $3/\phi$ which is the distance where the correlation drops below $0.05$.   

We incorporate the covariance function (\ref{eq:cov_fun}) into a probability model. Let $\calS = \{s_1,s_2,\ldots,s_n\}$ be the set of spatial locations. The $n\times 1$ vector $\omega$, whose $i$-th entry is $\omega(s_i)$, follows a multivariate normal distribution $N(0, K_{\theta})$, where $K_{\theta}$ is the $n\times n$ spatial covariance matrix with $(i,j)$th entry $K_{\theta}(s_i,s_j)$ in (\ref{eq:cov_fun}). The measurement errors are independent across locations, hence $\epsilon(s_i) \stackrel{iid}{\sim} N(0, \tau^2)$. This implies that the data vector $Z$, whose $i$-th element is $Z(s_i)$, is multivariate normal with mean vector $X\beta$, where $x(s_i)^{\top}$ are the rows of $X$, and variance-covariance matrix $K_{\theta} + \tau^2 I_n$, where $I_n$ denotes the $n \times n$ identity matrix. 

Spatial regression models, such as (\ref{eq:spatial_basic}), are fitted by estimating geostatistical parameters $\sigma^2$, $\phi$ and $\tau^2$ in addition to the regression coefficients $\beta$. We use (\ref{eq:spatial_basic}) to predict the outcome at a new location after accounting for the uncertainty in parameter estimates. When all points lie on a region represented as a 2-D plane, the distance between $s$ and $s'$ in (\ref{eq:cov_fun}) is given by the standard Euclidean distance formula. Here, the correlation drops at the same rate for every direction, so the spatial range is a function of distance only. Also, the covariance function in (\ref{eq:cov_fun}) ensures that $K_{\theta}$ is always positive definite \citep[see, e.g.,][]{banerjee}.

In our current context, the points lie along a curve representing the coastline. There are two issues. First, the Euclidean distance is inappropriate for modeling spatial covariances because the effective spatial range will be the distance \emph{along the coast} at which the correlation becomes negligible. Second, covariance functions that ensure positive definiteness in Euclidean coordinates need not be valid for other domains \citep{banerjee2}. This means that we will need to construct valid covariance functions along the coastline. Subsequently, we describe a simple approach to construct models such as (\ref{eq:spatial_basic}) using valid covariance functions for points along curves.  

%

\subsection*{Spatial processes for coastline measurements}\label{sec:spatial_process_coastline}

We now extend the model discussed in the previous section to the case where the data are observed over a coastline. Since all observations lie along the coastline, we will model spatial dependence along the coastline. The spatial range and variability will need to be interpreted in terms of distance along the coastline. Prediction is also sought at arbitrary points along the coast. 
We assume that any point $s$ on the coast is given by $\gamma(t)=(\gamma_1(t), \gamma_2(t))$ for some $t\in \calT \subset \Re^1$, where  $\gamma_1(t)=f(t)$ and $\gamma_2(t)=g(t)$ are parametric equations for the coordinates. Therefore, each value of $t$ determines a coordinate on a plane and traces out a curve $\gamma(t)$ as $t$ varies over a range $\calT$. The coastline is now given by the set of all points on it: $\gamma(\calT)=\{\gamma(t) : t \in \calT \subset \Re^1\}$. For example, a simple curve could be approximated by line segments. For each line segment, $\gamma(t)$ is a straight line, $\gamma_i(t)=\{s_i+t u \given t \in [0,\infty]\}$, originating at $s_i$ and parallel to the direction vector $u$. Here, $s_i= \begin {bmatrix} \gamma_{1i} \\ \gamma_{2i} \end{bmatrix}$, $u=\begin {bmatrix} u_1 \\ u_2 \end{bmatrix}$ and, hence, 
\begin{equation} \label{eq:line_seg} 
\gamma_i(t)= \left\{ \begin {bmatrix} \gamma_{1i}+tu_1 \\ \gamma_{2i}+t u_2 \end{bmatrix} \given t \in [0, \infty] \right\}\;. 
\end{equation}
A customary choice for the parameter $t$ is the arc length. As another example, consider a circular coast with radius $r$. The curve is defined as 
\begin{equation} \label{eq:arc_length} 
\gamma(t)=\{\gamma_1(t)=r\cos t, \gamma_2(t)=r\sin t \given t\in [0,\pi/2]\} 
\end{equation} 
The point $\gamma(t)=(r \cos t, r\sin t)$ moves in a fixed orientation (e.g., clockwise) as $t$ increases. If $t$ is the length of an arc of the circle and $\lambda$ is the angle in radians which the arc subtends at the center of the circle, then $t=r\lambda$. 

A spatial regression model such as (\ref{eq:spatial_basic}) can be defined over a coast by representing each point on the coast by $\gamma(t)$. Thus, we write $Y(t) = Z(\gamma(t))$ for every $t\in \calT$. Therefore,
\begin {equation} \label{eq:coastal_spatial} 
Y(t)=x^{\top}(t)\beta + \omega (t) +\epsilon (t);,
\end{equation}
where $x(t)$ is the vector of covariates observed at the point $\gamma(t)$, $\omega(t)$ is now defined over $\calT$ with covariance function 
\begin {equation}\label{eq:coastal_cov_fun} 
\mbox{Cov}(\omega(t), \omega(t'))=K_{\theta}(t,t')=\sigma^2 \exp(-\phi |t-t'|)\; ,
\end{equation}
where $|t-t'|$ is the absolute difference between $t$ and $t'$, and $\epsilon(t)\stackrel{iid}{\sim} N(0,\tau^2)$.

The choice of $t$ depends on the parametric equation used to approximate the coast. If the coast can be well-represented in closed form using a parametric equation, then $t$ as the arc-length is often a reasonable and convenient choice \citep[see, e.g.,][]{stewart_calculus}. The intuition stems from describing a curve by starting at a point on the curve and moving along the path traced by that curve. The point we arrive at after moving a distance of $t$ units on the curve is $\gamma(t)$. More generally, an arbitrary coastline can be well approximated using a series of small line segments. Each segment is then defined according to (\ref{eq:line_seg}). For example, in our subsequent simulation experiments we present linear approximations for an elliptical coastline. In our real example we use a series of small linear segments to model the coast along Waveland Beach in Mississippi.

\subsection*{Coastal kriging} \label{coastal kriging}
Exposure assessors may be interested in predicting the concentration of a toxicant at any arbitrary location on the coast. Let $Y(t_0)$ be the toxicant concentration measurement at the point $\gamma(t_0)$ on the coast. The posterior probability distribution of $Y(t_0)$, which is also referred to as the posterior predictive distribution, is computed in two steps. First, the unknown parameters in (\ref{eq:coastal_spatial}) are estimated by using Bayes' theorem to compute their posterior distributions. Thus, if $p(\theta,\beta, \tau^2)$ represents the prior distribution for unknown parameters and $p(y\given \theta,\beta, \tau^2)$ represents the likelihood, then the posterior distribution is given by
\begin{equation}\label{eq:bayes_theorem}
 p(\theta,\beta, \tau^2\given y) = \frac{p(\theta,\beta, \tau^2)\times p(y\given \beta,\theta,\tau^2)}{p(y)} \propto p(\theta,\beta, \tau^2)\times p(y\given\theta,\beta,\tau^2)\; .
\end{equation}
The prior distribution can be informative or non-informative. Non-informative priors typically deliver inference consistent with classical methods. Even for weakly informative priors, the inference is often close to classical methods because the effect of the data typically overwhelms the prior. While often producing inference numerically very similar to classical inference, Bayesian inference will retain simpler interpretability.   

Suppose we have toxin measurements at points $\gamma(t_1),\gamma(t_2),\ldots, \gamma(t_n)$ on the coast and have collected the $y(t_i)$'s in an $n \times 1$ vector $y$. Let $X$ be the $n \times p$ matrix with $i$-th row $x^{\top} (t_i)$ and $\omega$ be the $n\times 1$ vector with elements $\omega_i$. The posterior distribution of the model parameters is
\begin{align}\label{eq:posterior_full}
 p(\beta,\sigma^2,\tau^2,\phi\given y) & \propto U(\phi\given a_\phi, b_\phi) \times IG(\tau^2\given a_{\tau^2}, b_{\tau^2})\times IG(\sigma^2\given a_{\sigma^2}, b_{\sigma^2}) \nonumber \\ 
 &\qquad \times N(\beta \given \mu_\beta, V_\beta) \times N (y \given X \beta + \omega, \tau^2I)\;,
\end{align}
where $U(\cdot,\cdot)$, $IG(\cdot,\cdot)$ and $N(\cdot, \cdot)$ represent the uniform, the inverse-Gamma and the Normal distributions, respectively, as expounded in \citep{gelman}.

Posterior distributions, in general, are not available in simple closed-forms. Instead we sample $\{\beta, \omega, \theta,\tau^2\}$ from their posterior distribution, where $\theta=\{\sigma^2,\phi\}$, using Markov chain Monte Carlo (MCMC) methods \citep{gelman,banerjee}. Some simplifications are often made. One is to use a flat completely noninformative prior on $\beta$. Another is to integrate out $\omega$ from (\ref{eq:posterior_full}). The posterior samples for $\{\beta, \sigma^2,\tau^2,\phi\}$ are then obtained by simulating
\begin{align}\label{eq:posterior_collapsed}
p(\beta, \theta,\tau^2\given y) &\propto U(\phi\given a_\phi, b_\phi) \times IG(\tau^2\given a_{\tau^2}, b_{\tau^2})\times IG(\sigma^2\given a_{\sigma^2}, b_{\sigma^2}) \times N (y \given X \beta, K_{\theta}+\tau^2I)\;.
\end{align}  
The posterior samples for $\omega$ are subsequently obtained by sampling one instance of $\omega$ from $N(\cdot,\cdot)$ for each sampled value of $\{\beta, \sigma^2,\tau^2,\phi\}$. This is called composition sampling \citep{banerjee}. 

Suppose we have collected $M$ post-convergence posterior samples for the model parameters, say $\{\beta_{(j)}, \theta_{(j)},\tau^2_{(j)}\}$, for $j=1,2,\ldots,M$. Then the posterior samples for $Y(t_0)$ are obtained by composition sampling, i.e.,  for each $j$ we draw $Y_{(j)}(t_0)$  from the conditional normal distribution, say $N(m_{(j)}, v^2_{(j)})$, where the mean and variance are
\begin{align}
 m_{(j)}(t_0) &= x(t_0)^\top \beta_{(j)} +\tilde{K}_{\theta_{(j)}}(t_0,t) \tilde{K}_{\theta_{(j)}}^{-1}(t,t) \tilde{K}_{\theta_{(j)}}(t,t_0) (y-X\beta_{(j)} )\; \nonumber\\ 
 \mbox{ and }\; v_{(j)}^2(t_0) &= \tilde{K}_{\theta_{(j)}}(t_0,t_0)-\tilde{K}_{\theta_{(j)}}(t_0,t) \tilde{K}_{\theta_{(j)}}^{-1}(t,t) \tilde{K}_{\theta_{(j)}}(t,t_0) )\;, 
\end{align}
where $\tilde{K}_{\theta_{(j)}}(\cdot, \cdot)=K_{\theta_{(j)}}(\cdot, \cdot)+\tau^2 I$. Note that $m_{(j)}(t_0)$ and $v_{(j)}^2(t_0)$ are precisely the classical kriging estimator and variance evaluated at $\{\beta_{(j)}, \theta_{(j)},\tau^2_{(j)}\}$. Bayesian kriging, therefore, quantifies uncertainty in kriging by averaging the classical kriging estimator over the posterior distribution of the parameters. The resulting $Y_{(j)}(t_0)$ are samples from the posterior predictive distribution. The mean of these samples yields a point estimate of the predicted value at $t_0$, while the variance of the posterior samples estimates the predictive variance.  

One assumption to simplify matters is that $\phi$ and $\alpha=\frac{\tau^2}{\sigma^2}$ are fixed, say at values resulting from the empirical variogram \citep{banerjee}. Hence, the posterior samples for the model parameters are obtained from the conjugate model
\begin{align}\label{eq:posterior_simple}
p(\beta, \sigma^2 \given y) &\propto IG(\sigma^2\given a_{\sigma^2}, b_{\sigma^2}) \times N(\beta \given \mu_\beta, \sigma^2V_\beta) \times N (y \given X \beta, \sigma^2V_y)\;,
\end{align} 
where $V_y = R(\phi) + \alpha I$ and $R(\phi)$ is the spatial correlation matrix with elements $\exp(-\phi |t_i-t_j|)$. Here one can sample exactly from the posterior distribution in (\ref{eq:posterior_simple}). For each $j=1,2,\ldots,M$ we first draw $\sigma^2_{(j)} \sim IG(a^*_{(j)}, b^*_{(j)})$ followed by $\beta_{(j)}\given\sigma^2_{(j)},y \sim N(Bb,B\sigma^2_{(j)})$, where $a^*_{(j)}=a_{\sigma^2}+n/2$ and $b^*=b_{\sigma^2}+(y^\top V_y y-b^{\top} Bb)/2$, where $B=(X^{\top}V^{-1}_y X+V^{-1}_\beta)^{-1} $ and $b=X^{\top}V_y^{-1}y$. 

\section*{Simulation}\label{sec:sim}

The simulated data consists of $n=100$ data points. The outcome $Y(t)$ values were generated on an ellipse. We first generated $l_i \sim Uni(0, 2\pi) $ for $i=1,2,\ldots,n$, where the corresponding parametric equations are $m=2 \text{cos} (l)$ and $n=\text{sin}(l)$. We then drew a multivariate normal random variable $\omega \sim N(0,K_{\theta})$ and then $y(t_i) \sim N(\beta_0 + \omega(t_i), \tau^2)$, where $t_i$ is the arc-length between points $(m_{i-1}, n_{i-1})$ and $(m_i,n_i)$.

In the data generation step, we fixed $\tau^2=0.1$, $\beta=0$ and $\theta = \{1,1\}$. For assessing predictive performance, we used 75 observations for training the model and withheld 25 observations for testing the predictive validation.

We estimated the models in (\ref{eq:posterior_collapsed}) and (\ref{eq:posterior_simple}). To compare the performance of coastal kriging to kriging using Euclidean distance, we estimated the model in (\ref{eq:spatial_basic}) as well using the covariance in (\ref{eq:cov_fun}). For all models, we assigned a noninformative prior to $\beta_0$ (i.e., $V_{\beta}^{-1}=O$ the matrix of zeroes) and an $IG(2,2)$ prior to $\tau^2$. In (\ref{eq:posterior_collapsed}) $\sigma^2$ and $\phi$ were assigned $IG(2,2)$  and $U(0.8,30)$ priors. The $IG(2,b)$ prior provides a prior mean of $b$ but has, in theory, an infinite variance yielding a relatively vague prior but with a prior value centered around $b$. In (\ref{eq:posterior_simple}), we fixed $\phi = 1.07$ and $\alpha=0.25$ for the coastal kriging model and $\phi=22009.68$ and $\alpha=8.13 \times 10^{-5}$ for kriging with Euclidean distance. Starting values for $\sigma^2$, $\tau^2$ and $\phi$ in (\ref{eq:posterior_full}) and the fixed values for $\phi$ and $\alpha=\tau^2/\sigma^2$ in (\ref{eq:posterior_simple}) were provided using their estimates from the empirical variogram for the data \citep{banerjee}. 

We also compared coastal kriging to universal kriging (UK). Universal kriging is kriging with a trend, where $E(Z(s))$ is a linear combination of the known functions $\{f_0(s),\ldots, f_p(s)\}$ \citep{cressie}. We assume that the mean $E(Z(s))$ is a function of the coordinates in a linear form, i.e $Z(s)=\beta_0+\beta_1x_1(s)+\beta_2x_2(s)+\omega(s)+\epsilon(s)$, where $x_1(s)$ is the longitude at location $s$, $x_2(s)$ is the latitude at location $s$. In practice we will not have an exact parametric formula for the coastline. This needs to be approximated by simple parametric curves. The easiest such option is a sequence of line segments, as described earlier. We used our simulated dataset to evaluate the performance of such linear approximations. 

Let $\Delta m_i = m_i - m_{i-1}$ and $\Delta n = n_i - n_{i-1}$, then the length of the straight line segment connecting the two points is $t^*=\sqrt{(\Delta m)^2+(\Delta n)^2}$. For small $\Delta m$, the sum of the lengths of these line segments provides an approximation to the length of the curve. We will, therefore, consider four models for coastal kriging. The model in (\ref{eq:posterior_collapsed}) with the exact parametrization for an ellipse will be called Model~1a~, while that with linear approximation will be called Model~1b. Similarly, the exact and approximate parameterizations corresponding to the model in (\ref{eq:posterior_simple}) will be referred to as Model~2a~and Model~2b~respectively.

Table\ref{table: Model post sim} presents the posterior medians and $95\%$ Bayesian credible intervals for the parameters in each of the above four models, the simple Euclidean distance kriging model and the UK model. The credible intervals from all models include the true values of $\beta_0$. Models~1a~and~1b~captured the true values of $\sigma^2$ and $\phi$. Model~2b~also captured the true value of $\sigma^2$ and Models~2a~and~2b~captured the true value of $\tau^2$. To assess predictive performance across the six models, we used mean square prediction error (MSPE). Coastal kriging and UK models produced very similar MSPE values, and the highest MSPE was produced by the simple Euclidean distance kriging model. For model comparison we also used the Kullback-Leibler (K-L) divergence criterion ($D_{KL}(M_0 \given M_i)$), $i=1,\ldots 5$, where $M_0$ is the true distribution and $M_i$ is the distribution under model $i$. For multivariate normal distributions the Kullback-Leibler divergence (\cite{kl}) takes the form \begin{equation} \frac{1}{2} (tr(\Sigma^{-1}_{M_i} \Sigma_{M_0}) +[x \beta_{M_i}-x\beta_{M_0}]^\top \Sigma^{-1}_{M_i}[x \beta_{M_i}-x\beta_{M_0}]- n+ \text{ln} ( \text{det}(\Sigma_{M_i}))-\text{ln} (\text{det}(\Sigma_{M_0}))) \nonumber \end{equation} 
where $\Sigma=K_\theta + \tau^2 \text{I}$.
Model~1a~produced the lowest $D_{KL}$ followed by Models~1b,~2b~and~2a,~and the highest values were produced by the UK model and the simple Euclidean distance kriging model. We also used deviance information criterion (DIC), which is commonly used in Bayesian model selection. Model~2a~produced the lowest value followed by Models~2b,~1a~and~1b, then the UK model, and the highest value was produced by the simple Euclidean distance kriging model. Finally, ten-fold cross validation (CV(10)) was the lowest among coastal kriging models followed by the UK model then the simple Euclidean distance kriging model.

\begin{figure}[p]
\centering
\captionsetup{justification=centering,margin=2cm}
\includegraphics[width=\linewidth]{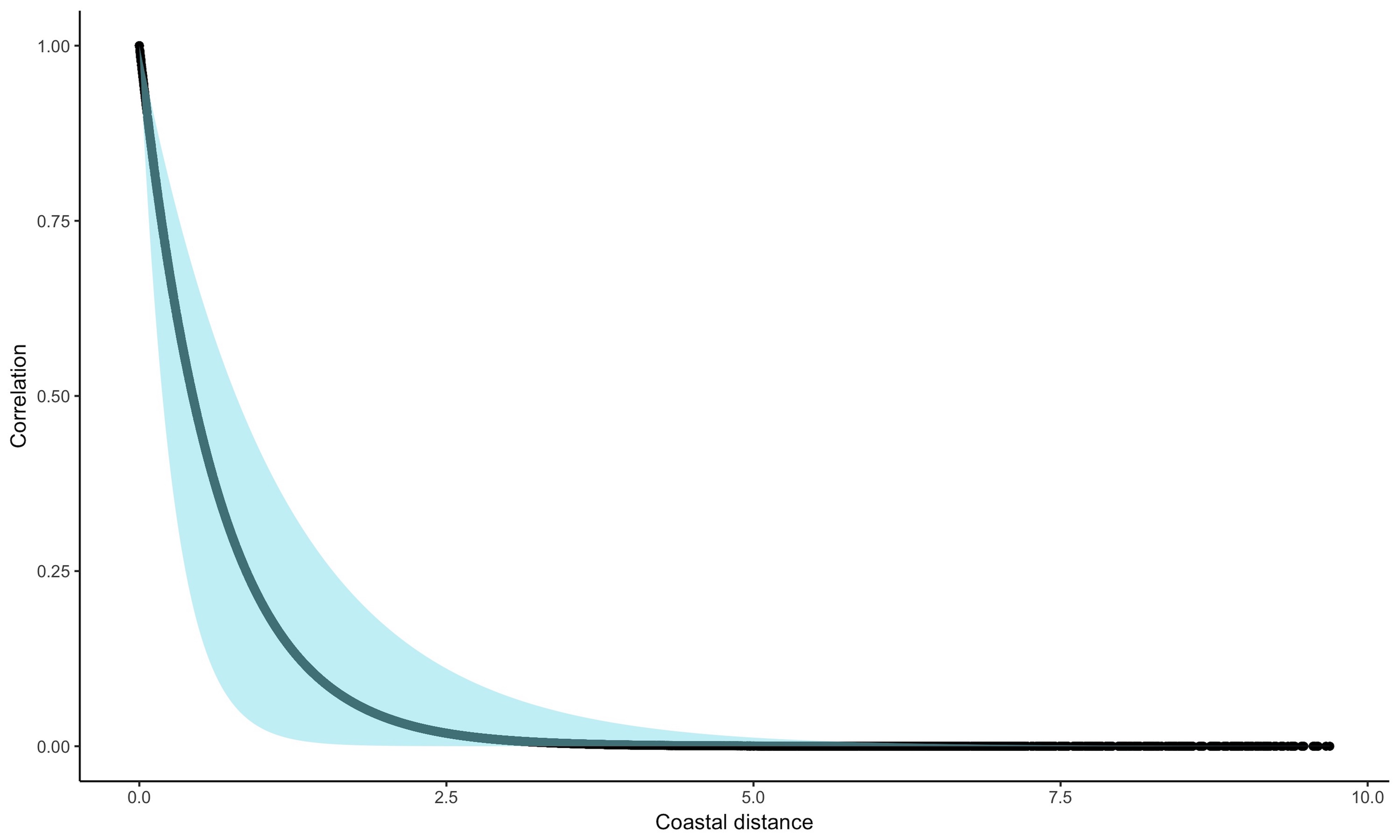}
\caption{Coastal kriging estimated correlation (solid line) versus coastal distance applying Model~1a to the simulated data along with 95\% credible bands (shaded).}
\label{fig:corr}
\end{figure}

\begin{figure}[p]
\centering
\captionsetup{justification=centering,margin=2cm}
\includegraphics[width=\linewidth]{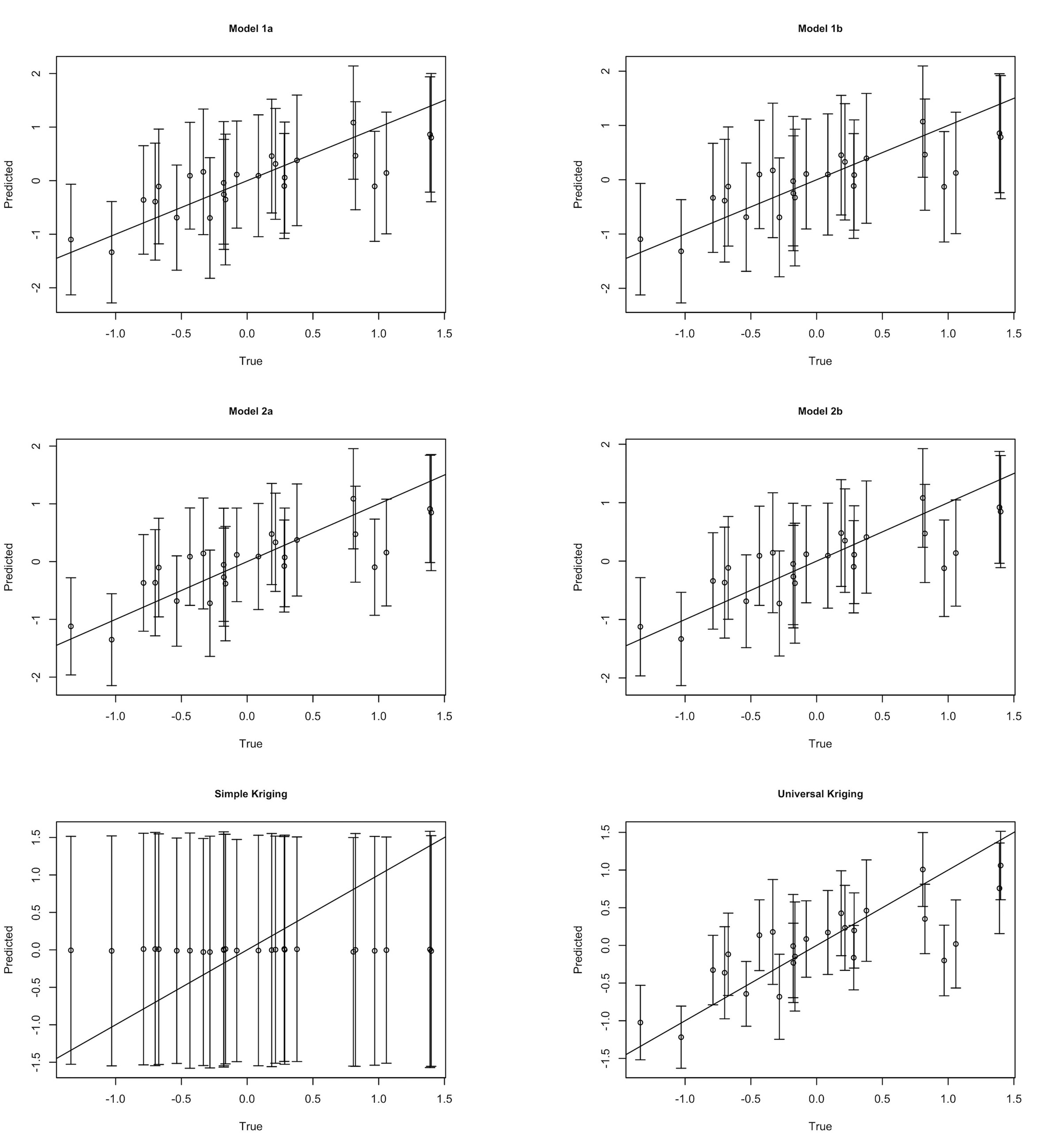}
\caption{Simulated data true versus predicted values with 95\% prediction intervals with $45^{\circ}$ line of the four coastal kriging and simple kriging models}
\label{Fig: T vs P sim}
\end{figure}

Figure~\ref{fig:corr} presents the ``coastal correlation'', i.e., the correlation function plotted against the distance along the coast. The solid line represents the posterior mean of $K_{\theta}(s,s')/\sigma^2$, while the shaded region represents the corresponding 95\% credible band providing uncertainty quantification for the spatial covariance using Model~1a. We also used  Bayesian $95\%$ prediction intervals and the predicted mean values of the outcome from the 25 holdout locations and plotted them against the true values; the results are seen in Figure~\ref{Fig: T vs P sim}. For the coastal kriging models, the intervals include the true values of the outcome variable in each of the holdout locations except for one location. The UK model provided improved prediction over simple Euclidean distance kriging model which produced the least accurate prediction with wider credible intervals.

These results indicate that Bayesian models using piecewise linear approximations to a parametric curve do not seem to adversely affect the inferential performance relative to models using the true form of the parametric curve. They also indicate that coastal kriging is better than classical kriging methods such as simple Euclidean distance kriging and UK when the source of variability in the data arises from a curve. Thus, Models~1b~and~2b~are good candidate models to be used in the data analysis.

\section*{Data Analysis}\label{sec:data}

\begin{figure} [p]
\centering
\captionsetup{justification=centering,margin=2cm}
\includegraphics[width=\linewidth]{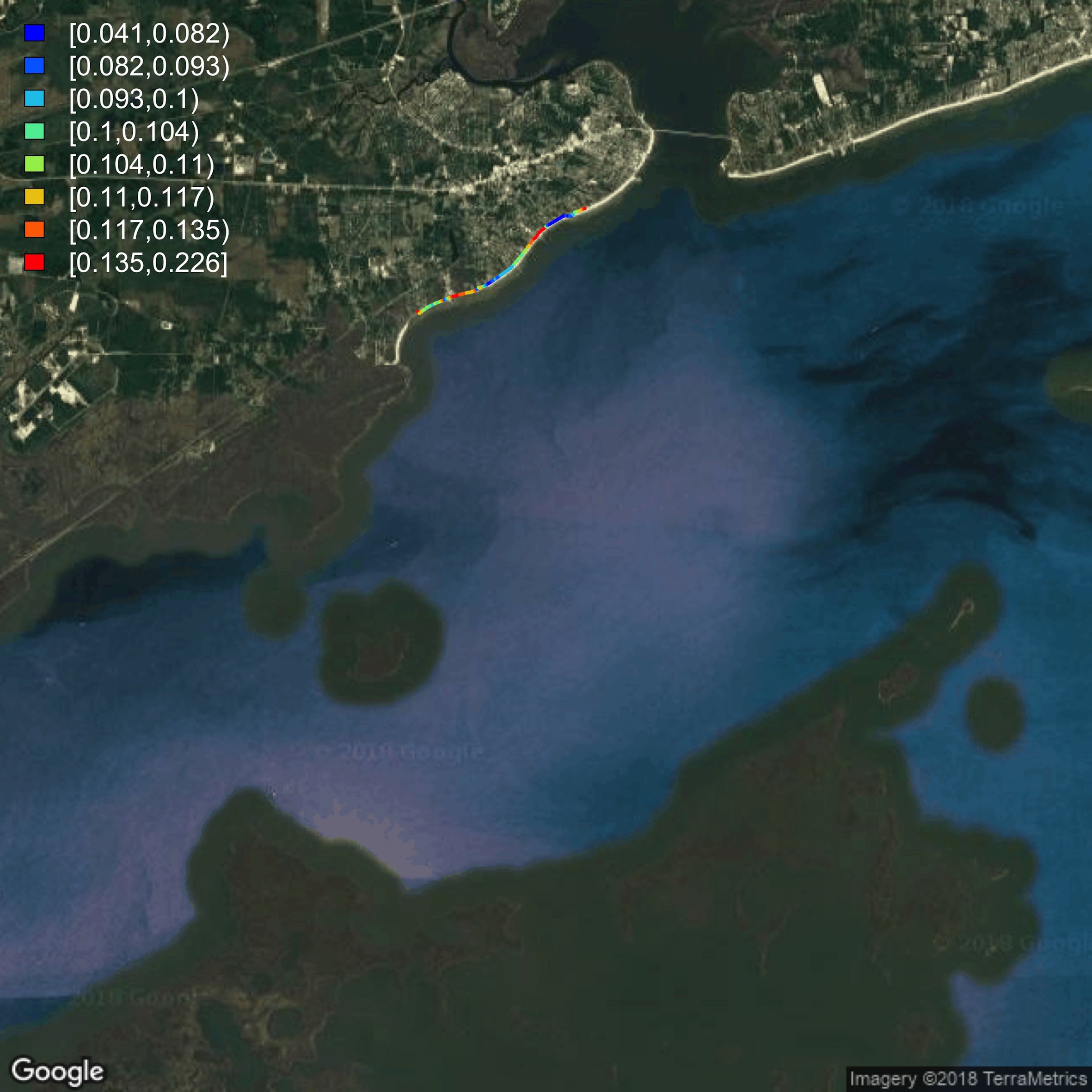}
 \caption{Map of observed and interpolated total hydrocarbons (ppm) from Model~1b over Waveland Beach, Mississippi} 
 \label{Fig:Data_map} 
 \end{figure}

Coastal kriging of the concentration of chemicals inhaled by the clean-up workers following the oil spill in 2010 may be useful to assess the potential health effects associated with the spill for locations without measurements. The data set used here consists of air samples collected on clean-up workers on Waveland beach, Mississippi which extends in an S-shape for seven or eight kilometers (Figure~\ref{Fig:Data_map}). The samples were collected for approximately 10 hours per day using passive dosimeters clipped to the workers' collars to measure breathing zone concentrations. The chemicals in the air diffused on to a charcoal pad inside the sampler. Five analytes were analyzed at the laboratory. They include total hydrocarbons (THC) which is a composite of the volatile chemicals in crude oil and is our main variable of interest. There were a total of 60 sample points (THC parts per million (ppm)) collected between September 19 and December 21, 2010 that were used in the analysis. Two exposure groups were considered, workers who cleaned jetties and other land-based structures and workers who cleaned beaches. 

Candidate models include Models~1b~and~2b~where the curve is approximated by line segments and the parameterization in (\ref{eq:line_seg}) is used. The fixed values of $\phi$ and $\alpha$ in (\ref{eq:posterior_simple}) could be the estimated from the variogram. However in coastal kriging, the variogram may not provide accurate estimates. Hence, we will use Model~1b~in the data analysis and compare the results to simple Euclidean distance kriging results. For both models, we assigned a noninformative prior to $\beta_0$ (i.e., $V_{\beta}^{-1}=O$ the matrix of zeroes) and an $IG(2,2)$ prior to $\tau^2$. In (\ref{eq:posterior_collapsed}) $\sigma^2$ and $\phi$ were assigned $IG(2,2)$  and $U(0.8,30)$ priors. The prior on $\phi$ implies that the effective spatial range, i.e., the distance beyond which spatial correlation is negligible, is between $0.1$ and $3.8$ on a coastline with a distance of $7.6$ kilometers. In addition, coastal kriging was compared to universal kriging (UK) with a linear trend.

Twelve observations acted as a holdout testing sample and the models were assessed based on their predictive performance at new locations using MSPE in addition to 10-fold cross validation (CV(10)) and on the goodness of fit measure DIC. All observations were log transformed to achieve normality. 

\begin{figure} [p]
\centering
\captionsetup{justification=centering,margin=2cm}
\includegraphics[width=\linewidth]{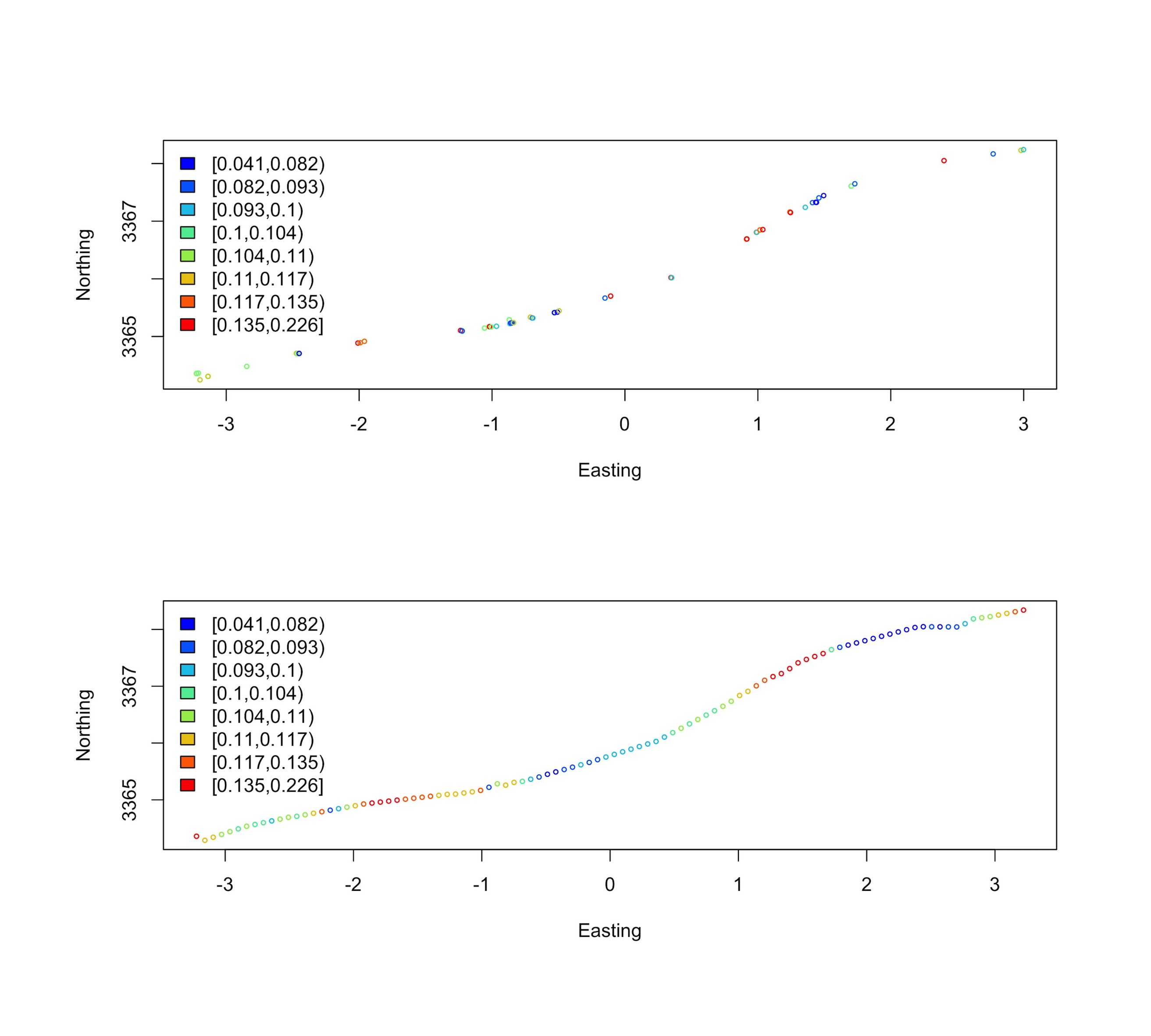}
 \caption{Observed total hydrocarbons (ppm) (top panel) and interpolated values from Model~1b (bottom panel) over Waveland Beach, Mississippi.} 
 \label{Fig:Data_inter} 
 \end{figure}

Table~\ref{table: model post data} shows parameter estimates of the fitted models. MSPE is almost the same among the three models, and the highest CV(10) resulted by the UK model. Model~1b~produced the lowest DIC value. Results show that coastal kriging proposed in (\ref{eq:coastal_spatial}) provides a better fit for coastal data compared to other classical kriging methods. The top panel of \ref{Fig:Data_inter} again shows the observed levels, while the bottom panel in \ref{Fig:Data_inter} shows interpolated total hydrocarbon (ppm) values obtained from Model~1b (posterior predictive means) along a string of 100 coordinates on the coastline. These figures evince the effect of coastal kriging: the interpolated values are in close agreement with the observations at locations close to those with observations, while smoothing the values at unobserved points by borrowing from neighboring observations. Figure~\ref{Fig:Data_map} shows a map of the observed \emph{and} interpolated measurements along the coastline overlaid on a \texttt{GoogleMap} with a legend indicating the level of the observed total hydrocarbons (ppm) over Waveland Beach, Mississippi. 

\section*{Discussion}\label{sec:disc}
We developed a simple and flexible Bayesian framework for spatially-oriented data that can be used to assess exposures of workers by interpolating levels of chemicals along a coastline. The statistical models for coastal kriging exploit a simple representation of the coast as a parametric function of the coordinates of points along the coastline. We presented four models using two different parameterizations. We found that for a simple curve, ``kriging'' using line segment approximation performs better than spatial kriging using Euclidean distance. This could be a useful and practical approach for kriging over any simple curve. The model is relatively easy to fit since the covariance depends on parameters in $\Re^1$.

We remark that the current article only considers worker exposure assessment, not community-based exposure assessment. In the GuLF STUDY  more than 28,000 samples of THC and several other chemicals were collected across the Gulf, along the coasts, and at ports and docks, providing sufficient data for the STUDY exposure estimates \citep{gulf}. These estimates were derived from groups of samples based on the tasks being performed.  The concentrations generated by these tasks (i.e. cleaning the beaches of oil and tar) represent task-derived exposures and, to a lesser extent, ambient air exposures.  Using such task-based measurements is not appropriate to impute general or community air concentrations because the task concentrations will be higher than ambient concentrations due to the workers being nearer to the source of the chemical emission than the community.  With the data used here, however, the imputed concentrations from the methodology described above may represent workers' exposures performing those same tasks in unmeasured locations. To date, occupational assessment methodologies have focused primary on fairly localized exposure situations. The method described here may be useful in more geographically extended situations, such as workers building a highway or mitigating a chemical release in a river or residents living along a fenceline adjacent to a manufacturing site.

Our study has some limitations within which our findings need to be interpreted carefully. First, the results are based on a total of 60 data points from which 48 were used in training the model and 12 were used in testing it. Second, the data points are distributed on a coast with little curvature which rendered the coastal kriging results slightly better than simple Euclidean distance kriging results. Last, but not least, the distribution of total hydrocarbons in the air is unknown and its source is not arising from the coast which may add some uncertainty in the fitted model, although in our data this uncertainty is assumed to be minimal.

Building valid models for coastal kriging presents many new research opportunities. For instance, it would be of interest to develop a model for more complicated coastlines, perhaps along closed curves such as the coasts of an island. 
Future work will also consider the modeling and analysis of censored data, as is commonplace in exposure studies, due to measurements below the limits of detection. Also, our current computations were cheap due to the relatively small number of spatial locations. Spatial models become expensive to compute for larger datasets, as perhaps would be the case with the full GuLF STUDY databases. Here, more specialized high-dimensional Bayesian models can be exploited \citep[see, e.g.,][]{banerjee2017high}. Finally, we will also consider extending this work to exposure assessment for communities rather than individuals.  
\newpage
\section*{Tables and Figures}

\begin{table}[h!]

\small
\caption{Medians, 2.5\% and 97.5\% quantiles of the posterior samples of the coefficient estimate, partial sill $\sigma^2$,
nugget effect $\tau^2$, decay parameter $\phi$, MSPE, DIC, Kullback-Leibler and CV(10) for the fitted models to the simulated data}
\begin{center}
\setlength\tabcolsep{2pt} 
\resizebox{\textwidth}{!}{\begin{tabular}{lccccccc}
  \hline 
\hline
 &True&  Model 1a$^1$ &Model 1b$^2$ & Model 2a$^3$&  Model 2b$^4$ & Simple kriging$^5$&Universal kriging  \\
\hline
$\beta_0$ &0 &0.16(-0.41,0.79)& 0.16 (-0.51,0.89) &  0.19(-0.35, 0.71)&0.24(-0.49, 0.97) &0.006(-0.16, 0.18)&0.03(-0.24,  0.31)\\
$\sigma^2$ &1 &0.55(0.30,1.09)&0.56 (0.30, 1.14)  & 0.48(0.36,0.67)&0.74(0.55, 1.04)& 0.58(0.43, 0.81)&0.18(0.12,0.26)\\
$\tau^2$&0.1 &0.18(0.12,0.28)&0.18 (0.12,0.27)& 0.12(0.09, 0.17) &0.12(0.09, 0.17)&$2.8\times 10^{-5}(2.1\times 10^{-5},3.9\times 10^{-5})$&0.17(0.12,0.25)\\
$\phi$ &1&1.20(0.85,2.81)&1.15 (0.71,3.58)&0.76 &0.76&31773.42 &0.32(0.16, 0.96)\\
MSPE& &0.57& 0.59 & 0.53 &0.54&1.23 &0.56\\
DIC & &30.11& 30.95 &28.80 &29.82 &55.43 &37.8 \\
Kullback-\\ Leibler & & 4.10& 5.61 & 5.68& 5.64& 73.67&100.7 \\
CV(10)&&0.170 &0.169&0.171&0.176&0.558&0.183\\
 \hline
   \hline
\end{tabular}
}
\end{center}
\caption*{
$^1$ Full hierarchical model using arc-length.\\
$^2$ Full hierarchical model using line segment approximation.\\
$^3$ Simplified hierarchical model using arc-length.\\ 
$^4$ Simplified hierarchical model using line segment approximation.\\ 
$^5$ Simplified hierarchical model using Euclidean distance.\\
}
\label{table: Model post sim}
\end{table}

\begin{table}[h!]

\small
\caption{Medians, 2.5\% and 97.5\% quantiles of the posterior samples of the coefficient estimate, partial sill $\sigma^2$,
nugget effect $\tau^2$, decay parameter $\phi$, and MSPE, DIC, and CV(10) for the fitted models of the log transformed total hydrocarbons}
\begin{center}
\begin{tabular}{lccccc}
  \hline 
\hline&Model 1$^1$ &Simple kriging$^2$ &Universal kriging\\
\hline
 $\beta_0$&-2.29(-2.71, -1.83)& -2.23(-2.67,-1.73) & -71.2(-8663.9, 7497.1) \\
$\sigma^2$ &0.59(0.29, 1.26)&0.59(0.28, 1.15) &0.52(0.34,0.89)  \\
$\tau^2$ &0.46(0.25, 0.80)& 0.46(0.27, 0.85)&0.17(0.12,0.23) \\
$\phi$&9.08(1.26, 24.82) &7.43(1.78, 22.70)&0.29(0.29,6.48) \\
MSPE  & 0.06& 0.06 &0.05\\
DIC & 34.4 & 38.6&65.05 \\
CV(10)&0.06&0.06&0.13\\
\hline  
   \hline
\end{tabular}
\end{center}
\caption*{$^1$ Full hierarchical model using line segment approximation.\\
$^2$ Full hierarchical model using Euclidean distance.\\
}

\label{table: model post data}
\end{table}

\newpage

%
%
\clearpage
\section*{Declaration of Publication}
Sudipto Banerjee and Gurumurthy Ramachandra acknowledge partial support from their grant CDC/NIOSH R01OH10093. Patricia Stewart, Mark Stenzel, Gurumurthy Ramachandran and Sudipto Banerjee have received funding from the NIH Common Fund and the Intramural Program of the NIH, National Institute of Environmental Sciences (ZO1 ES 102945) for other research on the Deepwater Horizon oil spill.  Some of the data generated under that funding were used in the example in the paper.
\clearpage
\bibliographystyle{biom}
\bibliography{mybib}
\clearpage

%

\end{document}